\documentclass[%
preprint,
 amsmath,amssymb,
 aps,
prl,
]{revtex4-1}

\usepackage{graphicx}
\usepackage{dcolumn}
\usepackage{bm}
\usepackage[mathlines]{lineno}

\usepackage[version=4]{mhchem}
\usepackage{amsmath, amssymb, amsthm}
\usepackage{mathtools}
\usepackage{multirow}

\newcommand{\vett}[1]{\mathbf{#1}}

\usepackage[utf8]{inputenc}
\usepackage[T1]{fontenc}
\usepackage{mathptmx}
\usepackage{etoolbox}

\makeatletter
\def\@email#1#2{%
 \endgroup
 \patchcmd{\titleblock@produce}
  {\frontmatter@RRAPformat}
  {\frontmatter@RRAPformat{\produce@RRAP{*#1\href{mailto:#2}{#2}}}\frontmatter@RRAPformat}
  {}{}
}%
\makeatother

\begin{document}

\preprint{APS/123-QED}

\title[]{Analytical Framework of Orbital Angular Momentum Beam under Misaligned Detection}
\author{Arttu Nieminen}
 \affiliation{ 
Tampere University, Faculty of Engineering and Natural Sciences, 33720 Tampere, Finland
}%
\author{Rizwana Ahmad}%
\author{Harald Haas}%
 \affiliation{ 
LiFi Research and Development Centre, Electrical Engineering Division, University of Cambridge, Cambridge, UK,}

\author{Humeyra Caglayan}
\email{h.caglayan@tue.nl}
 \affiliation{ 
Tampere University, Faculty of Engineering and Natural Sciences, 33720 Tampere, Finland
}%
\affiliation{%
Department of Electrical Engineering and Eindhoven Hendrik Casimir Institute,
Eindhoven University of Technology, Eindhoven 5600 MB, The Netherlands
}%

\date{\today}

\begin{abstract}

This work presents an analytical framework for modeling a detected orbital angular momentum (OAM) spectrum of an optical beam subject to tilt and lateral displacement. Firstly, we demonstrate that both types of misalignment generate OAM sidebands governed by the same functional form, each characterized by a distinct dimensionless parameter that depends on the beam's size and wavelength. Secondly, our analysis reveals that increasing the beam’s topological charge broadens the detected OAM spectrum. Lastly, we show that when both tilt and lateral displacement are present, the contribution of the original OAM mode can be tuned: specifically, by orienting the tilt and displacement in perpendicular directions, the resulting misalignment effects interfere destructively, thereby reducing crosstalk.
\end{abstract}

\maketitle

The concept of orbital angular momentum (OAM) of light was significantly advanced by the pioneering work of Allen et al. \cite{AllenOAM}, which demonstrated that light beams could possess OAM, characterized by a twisted wavefront. Allen's work laid the foundation for understanding how OAM can be utilized in various applications, such as optical communications \cite{wang2012terabit}, quantum information processing \cite{erhard2018twisted}, and microscopy \cite{microscopy}. Beams carrying OAM have the characteristic wavefront $e^{i\ell_0\phi}$, where $\phi$ is the azimuthal angle around the beam's propagation axis, and $\ell_0$ is an integer known as the topological charge of the beam, which describes the net change of phase in units of $2\pi$ in full round trip around the optical axis \cite{singularoptics}. The net OAM carried by a single photon is then given by $\hbar \ell_0$. 

There has been a particular rising interest in using OAM modes of light in optical communications \cite{oam_comm1,oam_comm2}: Light beams carrying different OAM are topologically protected from each other \cite{Willner:15}, giving the possibility of using the OAM modes as information carriers, effectively multiplying the amount of information transmitted \cite{wang2012terabit,OAMmultiplex1}. This OAM multiplexing is a specific type of mode division multiplexing (MDM) technique \cite{mdm}, where data is encoded into orthogonal modes of waveguide. When using OAM for multiplexing data, the OAM modes need to be effectively demultiplexed at the receiver, and it is an active field of research with numerous works using different techniques \cite{demultiplex_review}. 

However, detecting OAM modes presents unique challenges, particularly due to the detected OAM mode's sensitivity to wavefront errors caused by turbulence \cite{luo2022turbulence} and optical misalignments. Misalignments such as tilt and lateral displacement can distort the incoming light beam, leading to errors in the detection process \cite{Gibson:04}. Traditional OAM sorters, which are used to separate and identify these modes, have been reported to be difficult to align accurately \cite{oamsort}. Even minor deviations can result in significant signal degradation, affecting the overall performance and reliability of the communication system.

Understanding the effects of misalignment on OAM mode detection is crucial for developing robust and efficient optical communication systems. These findings are essential for designing encoding schemes that can mitigate the impact of misalignment errors, thereby reducing the error rate and enhancing the accuracy of data transmission. There have been works on modeling the effects of misalignment on the measured OAM analytically \cite{VasnetsovMV2005Aooa,quasiintrinsic,LiuOamCorrection} and correcting them experimentally \cite{zhao2017identifying}. The impact of random misalignment on the transmitted data rate on OAM multiplexed Radio-frequency (RF) waves has been studied \cite{millimeter-wave}. Furthermore, Lin et al. \cite{gaussianMisalign} studied Gaussian beams comprehensively, where the detected OAM distribution was shown to vary a lot depending on the direction of tilt and lateral displacement. 


To the best of our knowledge, a comprehensive and physically intuitive analysis of misalignment effects on optical OAM beams has been lacking. The most closely related work, by Vasnetsov et al. \cite{VasnetsovMV2005Aooa}, models misaligned beams as an infinite series of Bessel-Gauss modes. This mathematically elegant but conceptually opaque approach obscures the direct impact of tilt and displacement on the OAM spectrum. Understanding how tilt and lateral displacement affect the full optical power distribution is essential for designing and aligning OAM-based optical systems.

\begin{figure*}[ht!]
\centering
\includegraphics[width=1
\textwidth]{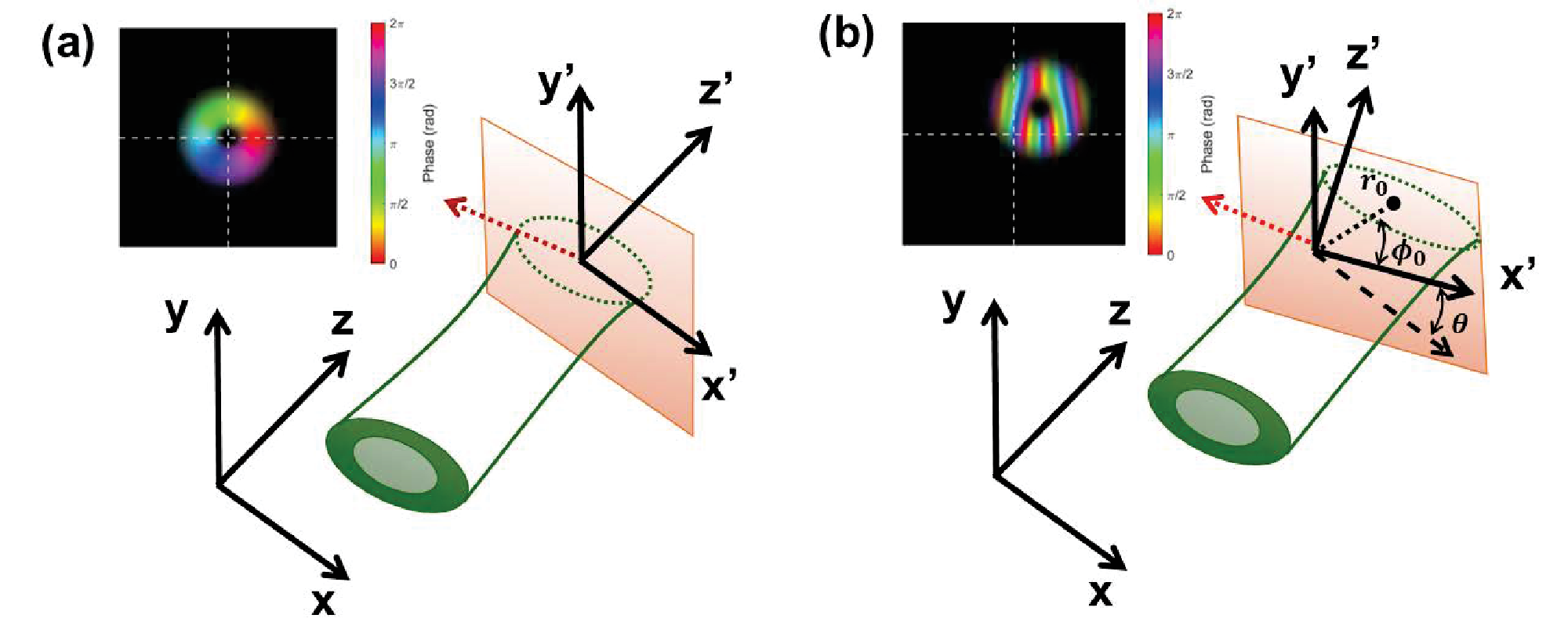}
    \caption{Schematic representation of an OAM beam in (a) reference frame which is aligned with its optical axis and origin, and (b) misaligned frame with tilt angle $\theta$ around y-axis and lateral displacement $r_0$ with polar angle $\phi_0$. Panel (a) shows the ring intensity profile and azimuthal phase profile being perfectly aligned with the origin. In panel (b), the beam center is not at the origin, and the phase profile shows a distorted phase profile due to the transverse wavevector caused by the tilt in the beam.} \label{misalignSchematic}
    \end{figure*}


Our work provides analytical solutions describing the effect of misalignment on OAM-carrying Laguerre-Gaussian (LG) beams, revealing the interplay between tilt and lateral displacement in the detected OAM spectrum. We show that when both misalignments occur along the same axis, they act independently, with no cross-effects. In contrast, when tilt and displacement have orthogonal components, their interaction becomes nontrivial—they can either amplify or suppress OAM mode crosstalk, depending on their relative orientation. Additionally, we link the sensitivity of the detected OAM distribution to the beam's topological charge, showing that higher-order modes experience increased error due to their larger spatial and Fourier-domain extent.

Throughout this work, we assume a uniformly polarised and paraxial optical beam propagating mainly in the z-direction: $E(x,y,z) = A(x,y,z)e^{ik_0z}$, where $k_0 = 2\pi/\lambda$ is the wavenumber, and $\lambda$ is the wavelength. The problem of detecting the complex amplitude of the beam in case of misalignment is presented in Fig. \ref{misalignSchematic}, where an optical beam is propagating in its on reference frame $\vett{r}=\{x,y,z\}$ while the detector plane is described in its reference frame $\vett{r'}=\{x',y',z'\}$, and it is assumed to lie in the plane $z'=0$. If the detector frame $\vett{r'}$ is perfectly aligned with the beam frame $\vett{r}$, the beam has the expected azimuthal phase profile in the middle of the detector, and this is illustrated in panel (a). In panel (b), the detector frame is shifted and tilted with respect to the optical axis of the beam. In this case, the beam seen at the detector is shifted away from the origin, and the phase profile of the beam is distorted from the azimuthal phase profile. We now consider a beam impinging onto a with lateral displacement $r_0$ in x-y-plane to polar angle given by $\phi_0$. Without loss of generality, we assume the beam to be tilted solely in the x-direction, corresponding to the detector plane being rotated around the y-axis by angle $\theta$. For further details, please refer to Sec. I of the supplementary material. In small tilt approximation,  the detected beam can be presented as follows:
\begin{equation}\label{eq:general}
E(x,y,z) = e^{i x k_0\theta}A(x - r_0\cos\phi_0,y - r_0\sin\phi_0,z)e^{ik_0z},
\end{equation}
where $A(x - r_0\cos\phi_0,y - r_0\sin\phi_0,z)$ represents the spatial envelope of the beam whose center is shifted by the lateral displacement. We now see the effect of tilt angle $\theta$ as an exponential term $e^{ixk_0 \theta }$, which causes x-directional tilt in the wavefront. The lateral displacement can be represented in a similar form by writing $A(x - r_0\cos\phi_0,y - r_0\sin\phi_0,z)e^{ik_0z}$ as a series of plane waves, i.e., in its angular spectrum representation \cite{mandel_wolf_1995}, so that the term is written in frequency domain polar coordinates $\{k_\perp,\varphi\}$ as $A(x - r_0\cos\phi_0,y - r_0\sin\phi_0,z)e^{ik_0 z} = \frac{1}{\sqrt{2\pi}}\int_{\vett{K}} \,e^{-ik_\perp r_0\cos(\varphi -\phi_0)} \tilde{A}(k_\perp, \varphi, z)\,e^{i \mathbf{K} \cdot \mathbf{R}}d^2K$, where $d^2K = k_\perp \, dk_\perp d\varphi$, and $\mathbf{K} \cdot \mathbf{R} = xk_\perp\cos\varphi + yk_\perp\sin\varphi$. The term in the integral $\tilde{A}(k_\perp, \varphi, z)$ is the angular spectrum of the beam, describing the amplitude of the plane wave component. Like this, lateral displacement is similar to tilt presented as exponential term $e^{-ik_\perp r_0\cos(\varphi -\phi_0)}$. This similarity is expected, as lateral displacement in the spatial domain corresponds to tilt in the Fourier domain by the translation property of Fourier transform \cite{goodman2005introduction}. 

Because the OAM modes of light come with the characteristic phase profile $e^{i\ell_0 \phi}$, it is more convenient to present them in cylindrical coordinates $E(\rho, \phi, z)$. In this coordinate system, we can expand the detected beam into a Fourier series of topological charges $E(\rho, \phi, z) = \frac{1}{\sqrt{2\pi}} \sum_{\ell=-\infty}^{\infty} a_{\ell}(\rho, z) e^{i\ell\phi}$, where the amplitudes are given as $a_{\ell} = \frac{1}{\sqrt{2 \pi}} \int_0^{2\pi} E(\rho, \phi, z) e^{-i \ell \phi} d\phi$. These Fourier series amplitudes indicate the power of OAM component $\ell$ at each radial distance $\rho$, and the total fraction of this OAM component is obtained by integrating its modulus squared over the whole radial space, namely $P_\ell = \int_0^\infty |a_{\ell}(\rho,z)|^2 \rho \, d\rho / \sum_{n} \int_0^\infty |a_{n}(\rho,z)|^2 \rho \, d\rho$, where the denominator normalizes the sum of powers to unity. \cite{oamcomponent} These relative OAM powers can be used to determine the full OAM spectrum of the detected beam. The OAM spectrum can also be defined similarly in its Fourier space using the angular spectrum given in Sec. II of the supplementary material.

In order to calculate the OAM content of the impinging light beam, we need to define its spatial profile. For this work, we use the Laguerre-Gaussian modes of light, which are the natural choice to describe the impinging paraxial beam, as they construct the full basis of solutions for the paraxial Helmholtz equation, and each mode carries an azimuthal index $\ell_0$, which also is the topological charge of the beam, i.e., its OAM. By knowing the dynamics of OAM detection for Laguerre-Gaussian modes, the response for any paraxial beam can be obtained. In the spatial domain, the envelope of Laguerre-Gaussian modes is given as \cite{Siegman}
\begin{align} \label{eq:LG_beam_maintext}
A^{LG}_{\ell_0,p}(\rho,\phi, z) = & \frac{C^{LG}_{\ell_0,p}}{w(z)} \left( \sqrt{\frac{2 \rho}{w(z)}} \right)^{|\ell_0|} L_p^{|\ell_0|} \left( \frac{2 \rho^2}{w(z)^2} \right) e^{-\frac{\rho^2}{w(z)^2}} \nonumber \\&\times  e^{i \frac{k \rho^2}{2 R(z)}} e^{-i \psi_{\ell_0,p}(z)}e^{i\ell_0\phi},
\end{align}
where $C^{LG}_{\ell_0,p} = \sqrt{\frac{2p!}{\pi (p+|\ell_0|)!}}$ is the normalisation constant, \( \psi_{p,\ell_0}(z) \) is the Gouy phase, and $L_{p}^{|\ell_0|}(x)$ is the generalised Laguerre polynomial. The beam size is determined by \( w(z) \), which is the beam waist at propagation distance $z$, and the wavefront curvature is determined by the radius \( R(z) \). The angular spectrum of the Laguerre-Gaussian beam is shown in the supporting information.

\section{OAM distribution when in the presence of either lateral displacement or tilt}

We now have all the necessary tools to calculate the OAM spectrum for misaligned Laguerre-Gaussian beams. For simplicity, we only consider the fundamental modes with radial index $p = 0$. We first start with the case where only one of the misalignments is present, i.e., only lateral displacement or tilt. A fully analytical solution can be obtained in this case, with the tilt case handled in the spatial domain and the lateral displacement in the Fourier domain. We find that the OAM spectrum in both lateral displacement and tilt case follow exactly the same formula when the beam profiles are inserted into their corresponding Fourier amplitudes, as shown in Sec. IV of the supplementary. $P_{\ell,\ell_0}(\Psi_\Omega)$ has the only difference in the definition of its adimensional parameter $\Psi_\Omega = \{\Psi_\xi,\Psi_\theta\}$, with $\Psi_\xi=r_0/w_0$ for the laterally displaced beam and $\Psi_\theta =k_0\theta w(z)/2$ for the tilted beam. The OAM formula can be compactly presented as follows:
\begin{equation} \label{OAM_one_misalign}
   P_{\ell,\ell_0}(\Psi_\Omega)=  \frac{(-1)^{|\ell_0|} \Psi_\Omega^{-2(|\ell_0|+1)}}{|\ell_0|!} \frac{\partial^{|\ell_0|}\left( \Psi_\Omega^2 e^{-\Psi_\xi^2} I_{|\Delta \ell|}(\Psi_\Omega^2)\right)}{\partial\left(1/\Psi_\Omega^2\right)^{|\ell_0|}},
\end{equation}
 where $\Delta\ell = \ell-\ell_0$. This formula is the first result of our work. It fully describes the dynamics of the external OAM of a LG beam when one of the misalignments is present in the system. The similarity in the formulas for the tilt and lateral displacement is expected by the aforementioned relation between the spatial and Fourier domain. To calculate the effect of misalignment for Laguerre-Gaussian beams with radial index $p>0$, the Laguerre polynomials can be presented as a finite sum of polynomial terms, which leads to a sum of similar differentiation formula as presented above, which is shown in Sec. V of the supplementary material.  The complete formulas are calculated by carrying out the differentiations, and they are a complicated combination of Bessel functions combined with polynomial terms, with the first few shown in Sec. VI of the supplementary material. 

The formulas indicate that to access the information on the OAM content of the OAM beam with topological charge $\ell_0$, we need to differentiate the function $\Psi^2 e^{-\Psi^2} I_{|\Delta \ell|}(\Psi^2)$ $|\ell_0|$ times with respect to $1/\Psi^2$. The $\Delta \ell$ is the distance between the original OAM mode and the probed OAM mode. This formula is independent of the sign of the topological charge $\ell_0$ and the probed OAM mode $\ell = \ell_0 + \Delta \ell$.

\begin{figure}[t]
\centering
\begin{tabular}{c c}
   \includegraphics[width=0.25\textwidth]{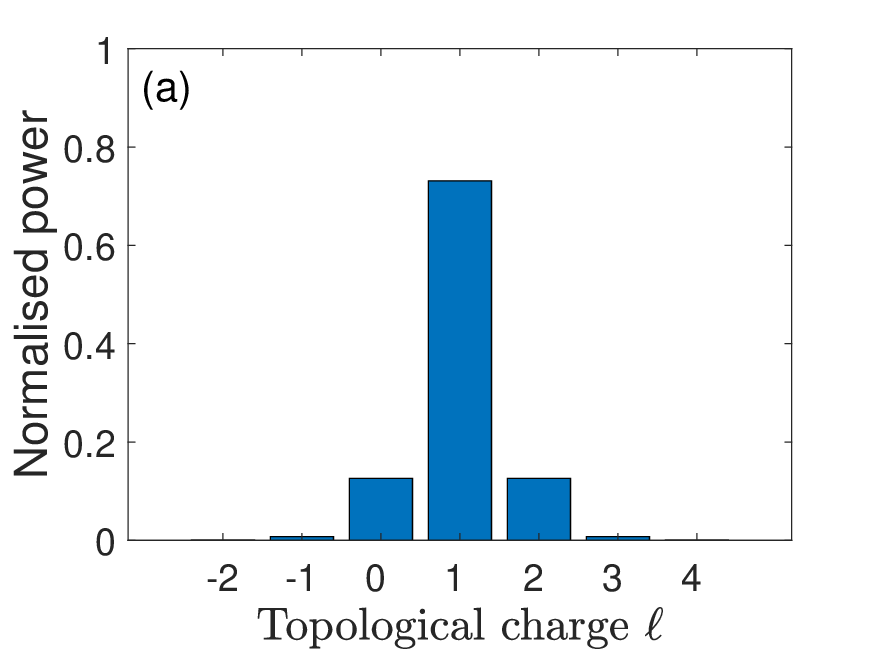}  &  \includegraphics[width=0.25\textwidth]{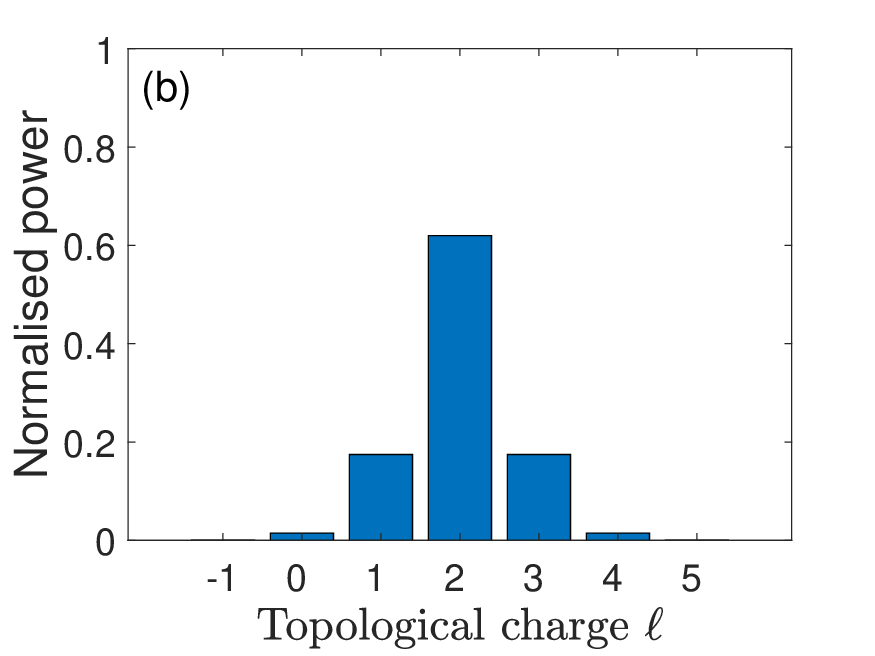} \\
   \includegraphics[width=0.25\textwidth]{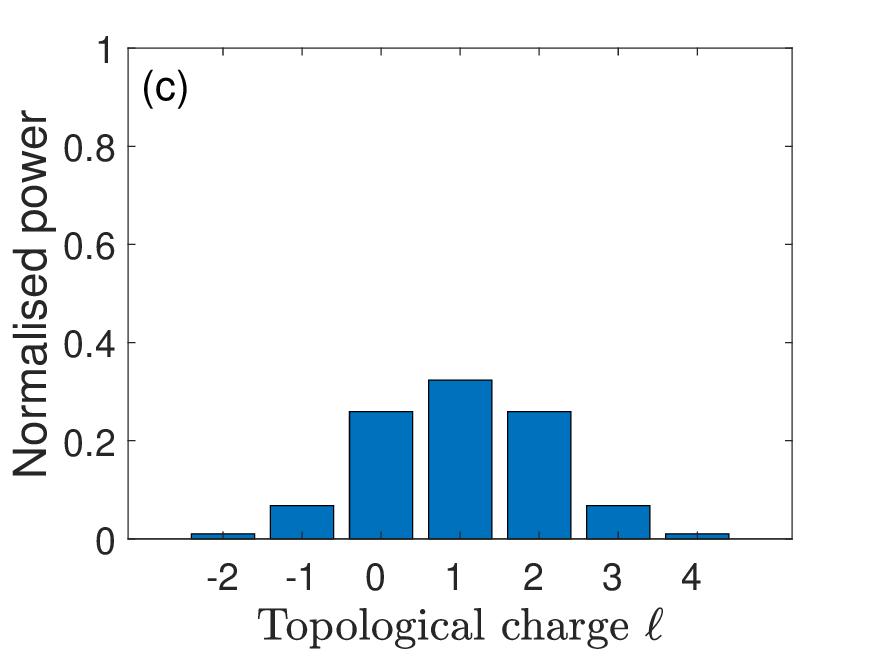}  &  \includegraphics[width=0.25\textwidth]{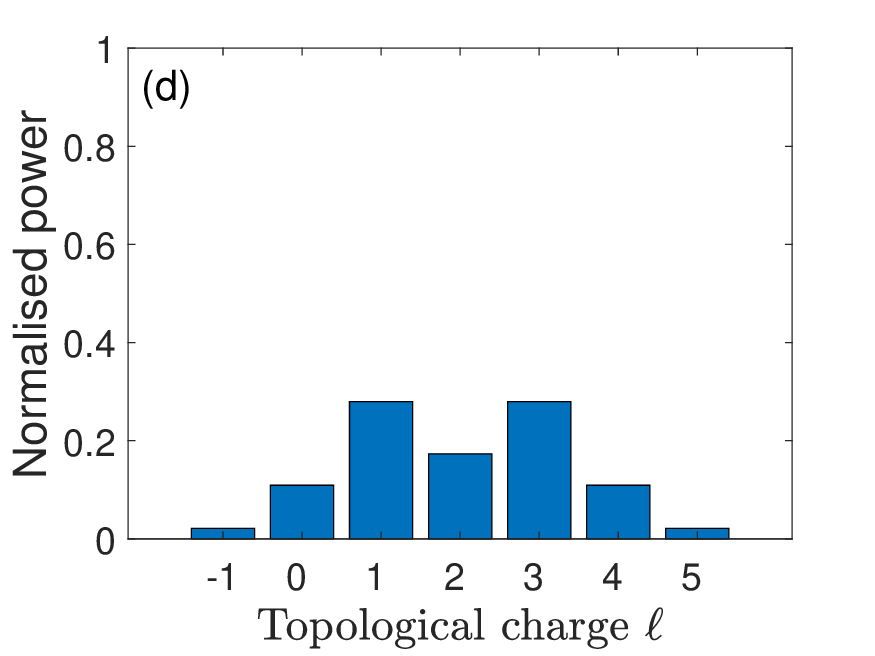}   
\end{tabular}
\caption{\footnotesize \label{misalignedModes1} OAM distribution for misaligned beam where (a) and (b) is for misalignments for OAM modes $\ell_0 = 1$ and $\ell_0 = 2$ with misalignment parameter $\Psi = 0.4$, and (c) and (d) correspond to the case when $\Psi = 0.8$. For 1 mm beam waist and 1550 nm wavelength beam, this corresponds to either displacement $ r_0 = 400$ $\mu m$ and $ r_0 = 800$ $\mu m$, or to tilt angle $\theta = 0.2$ mrad or $\theta = 0.4$ mrad.}
\end{figure}

An example of OAM distribution is shown in Fig. \ref{misalignedModes1}, where misalignment errors are shown with two different misalignment parameter $\Psi$ values, 0.4 and 0.8. The exact amount of lateral displacement or tilt depends on the wavelength and size of the impinging beam. For 1 mm beam waist and 1550 nm wavelength beam, this corresponds to either displacement $ r_0 = 400$ $\mu m$ and $ r_0 = 800$ $\mu m$, or to tilt angle $\theta = 0.2$ mrad or $\theta = 0.4$ mrad. The result shows how most of the energy is distributed to other OAM modes when the displacement is comparable to the beam waist of the beam, or when the tilt angle causes the transverse wavevector to be comparable to $2/w(z)$.

One of the immediate conclusions in the formulas is that tilt-induced and displacement-induced OAM errors exhibit an inverse dependence on the beam waist $w_0$. This suggests that if error tolerances for tilt and lateral displacement are known, we can find an optimal beam size at the detector with the minimum error amount in the detected OAM mode. To have a small error in the OAM detection, the lateral displacement $ r_0$ needs to be small with respect to the beam waist $w_0$. For tilt misalignment, the transverse wavevector caused by the tilt angle $\theta$ needs to be small with respect to $2/w_0$. We also can compare the tilt angle to the Gaussian beam divergence angle by writing $k_0w_0/2 = 1/\theta_0$, where $\theta_0$ is the cone half-angle, so that our tilt parameter becomes $\Psi_\theta = \theta/\theta_0$ at $z = 0$. We thus see that for the tilt misalignment to be small, the angle needs to be much smaller than the Gaussian beam divergence angle. This gives the physical explanation why larger beams and smaller wavelengths, which have lower divergence, are more sensitive to tilt than smaller beams and larger wavelengths.

Another important observation is that the OAM in lateral displacement is independent of the wavevector $k_0$, in contrast to the tilted case.
Moreover, the dependency of off-axis OAM on the beam waist $w_0$ remains unchanged even after the beam propagates over a distance $z$. The beam waist of the beam can be altered by manipulating its radius of curvature $R(z)$. In particular, if we have a propagated beam with beam width $w(z)$, we can introduce a collimating lens with focal length $f = R(z)$, which creates a new beam with beam waist $w(z)\mapsto w_0'$ \cite{fundamentalsofphotonics}. This will effectively reduce the OAM error in lateral displacement while keeping the tilt error unchanged. Using collimated beams is thus beneficial for minimizing OAM detection error in lateral misalignment.

\section{Discussion on effect of topological charge on OAM distribution}
 
We see in Fig \ref{misalignedModes1} that the OAM error is larger for $\ell_0 = 2$. To get a more intuitive understanding of the impact of the topological charge on the detected OAM distribution, we look at the OAM power on the original mode $|\ell_0|$ for small misalignments, such that Eq. \eqref{OAM_one_misalign} can be presented in its Taylor approximation in the quadratic order. In that case, all the OAM components are then at the original mode $\ell_0$ and the nearest neighboring modes $\ell_0 \pm 1$. In that case, the power in the $\ell=\ell_0$ mode is
\begin{equation} \label{taylorMisalign_dl0}
    P_{\ell_0,\ell_0}^{\Omega} \approx 1 - (1+|l_0|)\Psi_\Omega^2,
\end{equation}
which gives us topological charge dependent misalignment parameter $ \sqrt{1+|l_0| }\Psi_\Omega$, showing the decrease of the original OAM mode for increasing topological charge $\ell_0$. If we look at the modes $\Delta \ell = \pm1$ in the second-order polynomial approximations, we see that the energy is distributed to them evenly:
\begin{equation} \label{taylorMisalign_dl1}
    P_{\ell_0\pm 1,\ell_0 }^{\Omega}   \approx \frac{1}{2}(1+|l_0|)\Psi_{\Omega }^2,
\end{equation}
which tells that in small misalignment, the energy is first spread to the nearest neighbor modes according to $(1+|l_0|)\Psi_{\Omega }^2$. We will start seeing OAM power in $\Delta \ell = \pm2$ mode when $(\sqrt{1+|l_0|}\Psi_{\Omega })^4$ term becomes significant. By knowing the amount of misalignment in the beam, we can deduce how far from the original mode we have OAM sidebands.

The effect of misalignment scales with respect to the square root of $|\ell_0|$. This can be intuitively explained by the increased beam width and divergence for higher-order beams: the power of an OAM component is determined by radial integrals that are proportional to $(J_{\Delta \ell}(k_t \rho))^2$ and $(J_{\Delta \ell} (k_\perp r_0 ))^2$ for the tilt and lateral displacement cases, respectively.  These terms should be small over the beam's amplitude profile for misalignment to be small. In Ref. \cite{LGspotsize}, it was shown that the effective beam size and thus also the divergence angle of an LG beam scales as a function of topological charge increases as $w_{\ell_0}(z) =  w(z)\sqrt{1+\ell_0}$ and $ \theta_{\ell_0} =  \theta_0\sqrt{1+\ell_0}$, which scale exactly same as the misalignment term. The beam's transverse extent in the spatial domain is determined by its beam waist $w_{\ell_0}(z)$. In the Fourier space, analogous value can be written as a function of beam divergence angle $k_0\theta_{\ell_0}$. Thus, the increase in beam size and divergence angle can be attributed as the physical reason why the OAM error increases for both tilt and displacement cases for increasing $\ell_0$.

\section{Interaction of tilt and lateral displacement on OAM distribution}

\begin{figure*}[ht!]
\centering
\includegraphics[width=1
\textwidth]{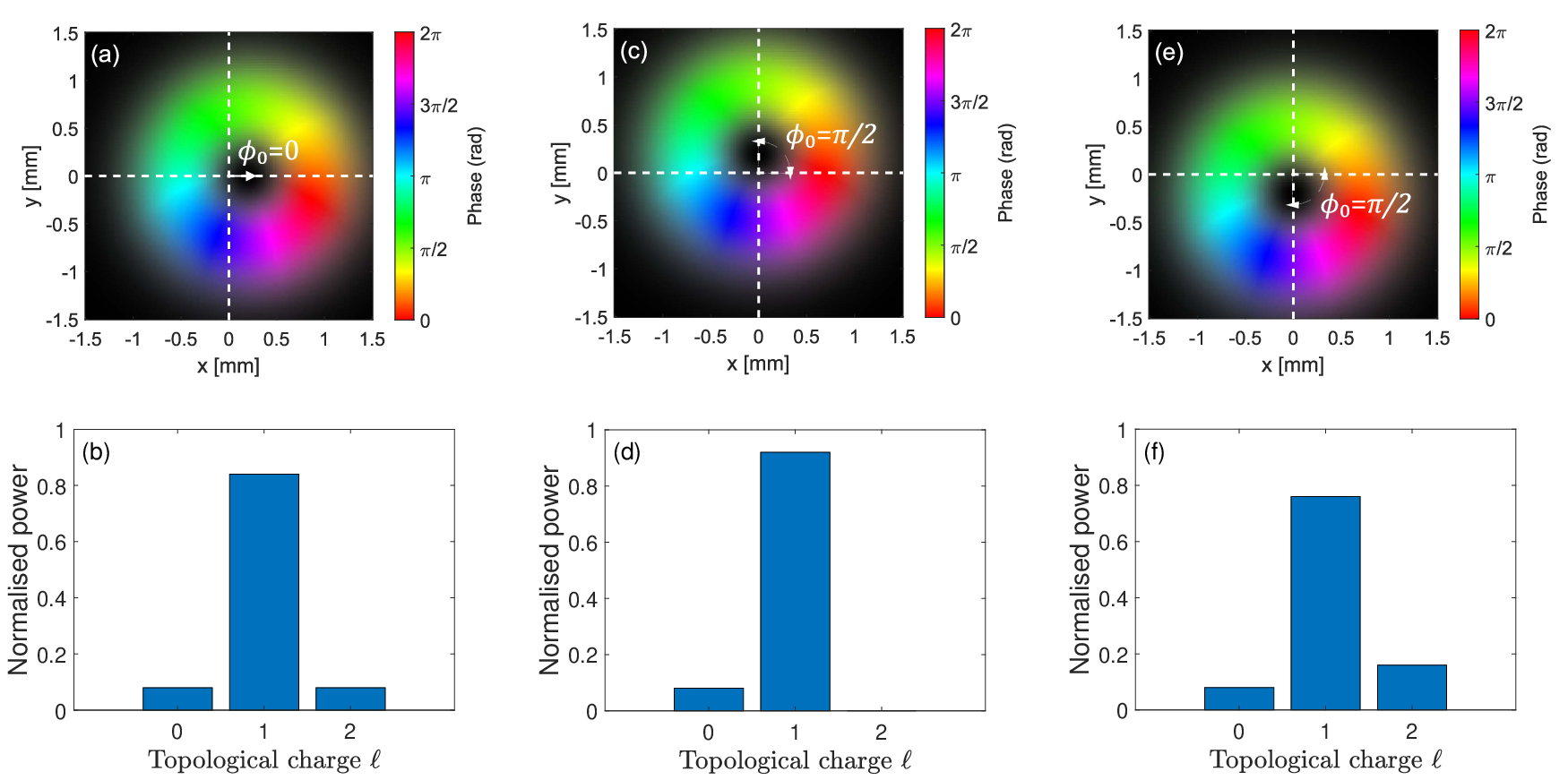}
    \caption{Beam profiles and OAM distributions for Laguerre-Gaussian mode $\ell_0 = 1$ with both misalignment parameters $\Psi_\xi = \Psi_\theta = 0.2$, corresponding to lateral displacement $x_0 = 200$ $\mu$m and tilt $\theta = 0.0057^o$. In all figures, the tilt is in the x-direction. Panels (a) and (b) depict the case where tilt and lateral displacement are in the same axis $\phi_{0} = 0$, (c) and (d) the case where beam is displaced upwards $\phi_{0} = \pi/2$, and in (e) and (f) the beam is displaced downwards $\phi_{0} = -\pi/2$. \label{bothMisalign}}
    \end{figure*}

We finally consider the case where both misalignments are present simultaneously, where we insert Eq. \eqref{eq:general} into the Fourier amplitude. The calculations are fully shown in Sec. VII of the supplementary material, where the resulting OAM spectrum formula a Taylor expansion with respect to our previously defined parameters $\Psi_\xi,\Psi_\theta$. In the supplementary, we also discuss the maximum order $N$, which can be considered sufficient when the maximum tolerance for misalignment is known. To understand the OAM distribution in different misalignments, we consider a specific case of the first fundamental LG mode $\ell_0 = 1$.  For simplicity, we only consider terms up to $2N = 2$, such that our OAM power distribution become quadratic polynomials $P_{\ell,1} = C^{\xi,\xi}_{\ell,1} \Psi_\xi^2 +  C^{\theta,\theta}_{\ell,1}  \Psi_\theta ^2 +C^{\xi,\theta}_{\ell,1} \Psi_\xi\Psi_\theta$, which in this case become
\begin{subequations}
    \begin{align}
        P_{1,1} = & 1 - 2 (\Psi_\xi^2 + \Psi_\theta^2) + 2 \Psi_\xi\Psi_\theta \sin\phi_{0} \\
        P_{2,1} = &  \Psi_\xi^2 + \Psi_\theta^2  - 2 \Psi_\xi\Psi_\theta\sin\phi_{0} \\
        P_{0,1} = &  \Psi_\xi^2 + \Psi_\theta^2,       
    \end{align} 
\end{subequations}
and for all other modes, $P_{\ell,1} = 0$. We immediately see that when either one of $\Psi_\xi,\Psi_\theta = 0$, our equations reduce into Eqs. \eqref{taylorMisalign_dl0} and \eqref{taylorMisalign_dl1}, as they should. Interesting dynamics can be observed depending on the value of $\phi_{0}$, i.e., the direction of the misalignments: When the two misalignments are on the same axis, so that $\phi_{0} = 0$, the OAM distribution becomes symmetric around $\ell_0 = 1$. However, when tilt and lateral displacement have perpendicular components, vastly different behaviour is observed: $\ell_0 + 1 = 2$ mode can either have increased or decreased power depending on the sign of $\sin\phi_{0}$, which is determined if the lateral displacement is in the positive or negative y-direction. We now consider the situation lateral displacement is completely in the y-axis, i.e., $\phi_{0} = \pm \pi/2$, such that it is perpendicular to tilt and that their contribution is the same $\Psi_\theta=\Psi_\xi \equiv \Psi$. In this case, we have $P_{1,1} = 0$ for $\phi_{0} = \pi/2$, and $P_{1,1} = 4\Psi^4$ for $\phi_{0} = -\pi/2$. The energy of $P_{1,1}$ comes from the original OAM mode $\ell_0 = 1$, while $P_{0,1}$ stays completely unaffected on the misalignment directions. These dynamics are depicted in Fig. \ref{bothMisalign}, which shows the beam profile and corresponding OAM distributions for small misalignment values $\Psi_\xi=\Psi_\theta = 0.2$.

Interesting behaviour is observed for higher topological charges: For $\ell_0 >1$, both $\Delta \ell \pm1$ components have $\Psi_\xi\Psi_\theta\sin\phi_{0}$ cross terms, with $P_{\ell_0+1,\ell_0}$ having $(\ell_0+1)\Psi_\xi\Psi_\theta\sin\phi_{0}$ and $P_{\ell_0-1,\ell_0}$ having $(\ell_0-1)\Psi_\xi\Psi_\theta\sin\phi_{0}$. This even more strongly focuses all OAM power to $\ell=\ell_0$ for $\phi_{0}=\pi/2$, and distributes OAM power to neighbouring modes for $\phi_{0}=-\pi/2$, makes the OAM distribution more symmetric than for $\ell_0 = 1$ case.  Sec VIII of Supplementary Information discusses OAM distributions for higher topological charge $\ell_0 > 1$. Fig. S1 shows this effect for $\ell_0 = 10$. 

In conclusion, we present a comprehensive analytical framework for understanding how misalignment— tilt and lateral displacement—affects the detection of orbital angular momentum (OAM) modes in optical beams. We show that these effects are particularly pronounced when the misalignments are orthogonal, leading to constructive or destructive interference in the detected OAM spectrum. Crucially, we link the increased susceptibility of higher-order OAM modes to their broader spatial and angular extent, which tightens alignment tolerances. This insight enables the prediction and control of OAM detection errors as a function of topological charge—an essential step toward designing more robust and precise OAM-based optical communication systems. Our results offer a practical foundation for minimizing misalignment and advancing the reliability of future OAM-enabled technologies.



%
%

%

\begin{acknowledgments}
A. N. thanks Marco Ornigotti and Robert Fickler for fruitful discussions and valuable comments. This work was funded by the Research Council of Finland under the CHIST-ERA (CHIST-ERA-21-NOEMS-002) grant number 357746, and Economic and Social Research Council (ESRC) under the CHIST-ERA grant number EP/X034542/2, titled "MEMS-Metasurface Based Tunable Optical Vortex Lasers for Smart Free-Space Communication (Meta-LiFi)".
\end{acknowledgments}

\section*{Data Availability Statement} 
The data that support the findings of this study are openly available in [repository name, e.g., “figshare”] at http://doi.org/[doi], reference number [reference number].

\appendix

\bibliography{references}

\end{document}